\documentclass[prd,reprint,amsmath,amssymb,floatfix,aps,longbibliography,superscriptaddress,nofootinbib]{revtex4-2}

\usepackage[normalem]{ulem}

\usepackage{amsmath,amssymb,amsthm,bm,braket,ascmac,bbm}
\usepackage{graphicx}
\usepackage[colorlinks=true, allcolors=blue]{hyperref}
\usepackage{slashed}
\usepackage{enumerate}
\usepackage{enumitem}

\usepackage[normalem]{ulem}
\usepackage{soul}

\setstcolor{red}

\begin{document}

\title{Wiggly dilaton: a landscape of spontaneously broken scale invariance
}

\author{Sudhakantha Girmohanta}
\email{sgirmohanta@sjtu.edu.cn}
\affiliation{Tsung-Dao Lee Institute, Shanghai Jiao Tong University, \\
No.~1 Lisuo Road, Pudong New Area, Shanghai, 201210, China}
\affiliation{School of Physics and Astronomy, Shanghai Jiao Tong University, \\
800 Dongchuan Road, Shanghai, 200240, China}

\author{Yuichiro Nakai}
\email{ynakai@sjtu.edu.cn}
\affiliation{Tsung-Dao Lee Institute, Shanghai Jiao Tong University, \\
No.~1 Lisuo Road, Pudong New Area, Shanghai, 201210, China}
\affiliation{School of Physics and Astronomy, Shanghai Jiao Tong University, \\
800 Dongchuan Road, Shanghai, 200240, China}

\author{Yu-Cheng Qiu}
\email{ethanqiu@sjtu.edu.cn}
\affiliation{Tsung-Dao Lee Institute, Shanghai Jiao Tong University, \\
No.~1 Lisuo Road, Pudong New Area, Shanghai, 201210, China}
\affiliation{School of Physics and Astronomy, Shanghai Jiao Tong University, \\
800 Dongchuan Road, Shanghai, 200240, China}

\author{Zhihao Zhang}
\email{zhangzhh@sjtu.edu.cn}
\affiliation{Tsung-Dao Lee Institute, Shanghai Jiao Tong University, \\
No.~1 Lisuo Road, Pudong New Area, Shanghai, 201210, China}
\affiliation{School of Physics and Astronomy, Shanghai Jiao Tong University, \\
800 Dongchuan Road, Shanghai, 200240, China}

\date{\today}

\begin{abstract}
The dilaton emerges as a pseudo-Nambu-Goldstone boson (pNGB) associated with the spontaneous breaking of scale invariance in a nearly conformal field theory (CFT).
We show the existence of a wiggly dilaton potential that contains multiple vacuum solutions in a five-dimensional (5D) holographic formulation.
The wiggly feature originates from boundary potentials of a 5D axion-like scalar field,
whose naturally small bulk mass parameter corresponds to a marginally-relevant deformation of the dual CFT.
Depending on the energy density of a boundary $3$-brane,
our model can provide a relaxion potential or generate a light dilaton. However, an extremely light dilaton requires fine-tuning.
\end{abstract}

\maketitle

\section{Introduction}

Spontaneous breaking of approximate scale invariance is an attractive idea
which, for instance, may offer a solution to the electroweak naturalness problem in the Standard Model
by dynamically generating the electroweak scale hierarchically smaller than the Planck scale.
Associated with spontaneously broken scale invariance (SBSI),
a pseudo-Nambu-Goldstone boson (pNGB), called {\it dilaton}, emerges
and it provides rich phenomenology and cosmology
(see e.g. refs.~\cite{Csaki:2000zn,Csaki:2007ns,Abu-Ajamieh:2017khi,Girmohanta:2023tdr}).
To discuss dilaton physics, its potential shape is particularly important because it governs the dynamics of the dilaton,
which has a significant impact on the evolution of the Universe,
as well as the vacuum expectation value (VEV), controlling the generated mass scale,
and the dilaton mass.

According to the AdS/CFT correspondence~\cite{Maldacena:1997re,Gubser:1998bc,Witten:1998qj} (see also refs.~\cite{Arkani-Hamed:2000ijo,Rattazzi:2000hs}), a concrete realization of a four-dimensional (4D) nearly conformal field theory (CFT) with SBSI is provided by the Randall-Sundrum (RS) model equipped with a compact extra dimension
\cite{Randall:1999ee} where two 3-branes, named UV and IR branes, reside on two fixed points.
The bulk geometry is warped, which rescales physical quantities on the IR brane by a warp factor,
exponentially depending on the size of the extra dimension (or the distance between the two $3$-branes).
The existence of the IR brane corresponds to SBSI in the dual 4D CFT picture.
The distance between the two $3$-branes is described by a {\it radion} degree of freedom, which is
identified as the dilaton.
Therefore, in the five-dimensional (5D) model, the stabilization of the radion by providing its potential is a critical issue. 

The Goldberger-Wise (GW) mechanism
\cite{Goldberger:1999uk} has been a classic example that exploits a 5D scalar field with a nontrivial bulk profile
to create a potential to stabilize the radion.~\footnote{
For other radion stabilization mechanisms, see e.g. ref.~\cite{Fujikura:2019oyi} and references therein.} 
To generate an exponential hierarchy of energy scales between the UV and IR branes,
a bulk mass parameter of the 5D scalar field must be suppressed by a parameter $\epsilon \ll 1$,
which corresponds to a small marginally-relevant deformation of the dual CFT, $\Delta \mathcal{L} = g \mathcal{O}_g$
where ${\rm Dim}[\mathcal{O}_g] = 4 - \epsilon$,
to trigger SBSI after a long period of the renormalization running.
However, to the best of our knowledge, such a small $\epsilon$ is always assumed and
there has been no discussion to justify the assumption.
Moreover, boundary values of the 5D scalar field are usually fixed by taking infinitely large couplings of boundary potentials
for simplicity of calculations, which are totally artificial.~\footnote{
One exception is given in ref.~\cite{Girmohanta:2024kyx}.
} 

\begin{figure}
    \centering
    \includegraphics[width=7.5cm]{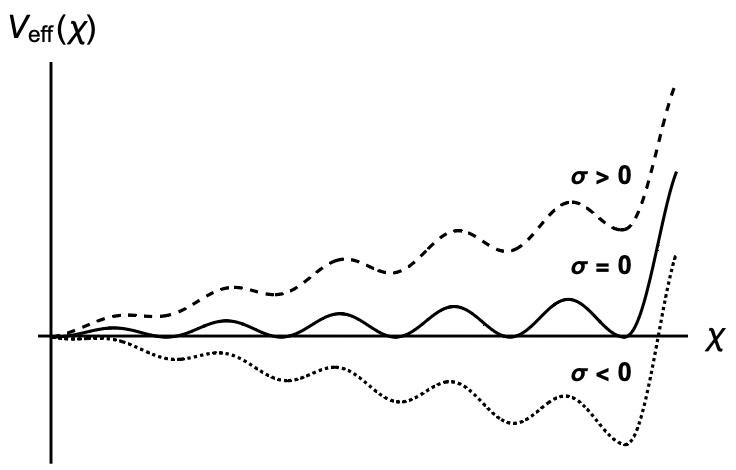}
    \caption{Sketch of the dilaton/radion effective potential $V_{\rm eff} (\chi)$ in three different scenarios of the IR brane tension.
    Here, $\sigma$ stands for a perturbation in the IR brane tension defined in the main text.
    The dilaton potential has multiple locally stable vacuum solutions.}
    \label{fig:main}
\end{figure}

In the present paper, we explore a possibility that the 5D scalar in the GW mechanism is given by an axion-like field (ALP)
whose pNGB nature suppresses its bulk mass term.
Periodic potentials of the axion field at boundary branes
are considered to be dynamically generated as in the case of the QCD axion, and they are naturally smaller than brane tensions.
We then find a nontrivial effective dilaton/radion potential, which wiggles and has multiple local minima as sketched in Fig.~\ref{fig:main}. 
When we parametrize a perturbation of the IR brane tension (denoted by $\sigma$) as $\xi = -\Lambda_5 / k |\Lambda_1| = 1+\sigma$ where $\xi$ gives the relative strength between the bulk cosmological constant $\Lambda_5<0$ and the IR brane tension $\Lambda_1<0$.
$k$ is the curvature scale.
Depending on the sign of the perturbation $\sigma$, the obtained effective dilaton potential behaves differently.
For $\sigma<0$, the potential $V_{\rm eff} (\chi)$ of dilaton/radion $\chi$ can have a global nontrivial minimum.
As we will see, this case could achieve the dilaton with a parametrically small mass,
$m_\chi \propto \epsilon$,
while the conventional study on the GW mechanism gives $m_\chi\propto \sqrt{\epsilon}$~\cite{Coradeschi:2013gda}.
On the other hand, for $\sigma>0$, the dilaton potential is overall an increasing function of $\chi$, which may be able to serve as a realization of a relaxion model~\cite{Dvali:2003br,Graham:2015cka}.~\footnote{The relaxion scenario can be realized by an axion-like field~\cite{Gupta:2015uea,Choi:2015fiu}.
Ref.~\cite{Fonseca:2017crh} considers an extra dimension setup where a tilted relaxion potential is provided directly by a 5D axion field and the size of the extra dimension is the same in each local minimum. Hence, the setup is different from ours.
Ref.~\cite{Hamada:2020bbf} also constructs a relaxion model in a different 5D geometry with one $3$-brane.}

The rest of the paper is organized as follows.
Section~\ref{sec:model} presents our 5D setup for radion stabilization introducing a 5D axion field
with periodic boundary potentials.
Then, in section~\ref{sec:VEVs}, we show the existence of multiple vacuum solutions.
Finally, section~\ref{sec:summary} is devoted to conclusions and discussions.

\section{Setup}
\label{sec:model}

Let us consider a 5D spacetime, labeled by $(x^\mu,y)$ with $y$-direction compactified, whose topology is $S^1/\mathbf{Z}_2$.
Two $3$-branes are located on two orbifold fixed points whose distance describes the size of the extra dimension.
We stabilize the distance by a 5D axion-like field $a$ whose mass is suppressed by a small dimensionless parameter $\epsilon$.
To be specific, we consider that $a$ arises from a complex phase mode of a bulk scalar field $\phi$ that is charged under a spontaneously broken global $U(1)$ symmetry.
~\footnote{We consider that the radial mode of $\phi$ is heavy. So it has a constant solution that can be neglected and does not impact the background geometry.}
The mass of the axion $a$ comes from a small explicit breaking of the global symmetry. 
The quality of the axion could be guaranteed by making this global symmetry accidental~\cite{Izawa:2002qk, DiLuzio:2017tjx, Lillard:2017cwx, Lee:2018yak, Lillard:2018fdt, Cox:2019rro, Ardu:2020qmo, Nakai:2021nyf, Lee:2021slp, Choi:2022fha, Qiu:2023los, Chaffey:2023xmz, Nakagawa:2023shi}.
To simplify the discussion, we consider that $\phi$ is subjected to a gauged $\mathbf{Z}_N$ symmetry. Then the leading order $U(1)$ explicit breaking operator is $\propto \lambda \phi^N$, where $|\lambda|\sim \mathcal{O}(1)$, and the mass of $a$ is generated from it.
For a large $N$, this explicit breaking can be suppressed to a satisfying level, and small $\epsilon$ is justified.
For simplicity, we work in natural units, $ c=\hbar=1$, considering that $a$ is dimensionless.
The 5D action is 
\begin{align}
    S & = \int d^4x dy \sqrt{g} \left( -\frac{R}{2\kappa^2}  + \frac{1}{2\kappa^2} g^{MN}\partial_M a \partial_N a - V(a) \right) \nonumber \\
    & \quad - \int d^4 x \sqrt{g_0} \, V_0(a) - \int d^4 x \sqrt{g_1} \, V_1(a) \;, \label{eq:action}
\end{align}
where $\kappa^2 \equiv 1/(2M_*^3)$ with the 5D fundamental scale $M_*$,
$R$ denotes the 5D Ricci scalar describing the 5D curvature,
$\sqrt{g} = \sqrt{|{\rm det} g_{MN}|}$ and $g^{MN}$ ($M,N = 0,1,2,3,5$) is the inverse metric.
The bulk potential is given by $V(a)$. Here the brane potential located at $y_i$ is labeled as $V_i$,
where $i=0,1$ respectively corresponds to the UV and IR branes.
At the location $y_i$, $\sqrt{g_i} = \sqrt{|{\rm det}g_{\mu\nu}(y_i)|}$ denotes the metric determinant of $g_{\mu\nu}$
with $\mu,\nu=0,1,2,3$.
The ansatz solution for the metric $g_{MN}$ is~\cite{Randall:1999ee} 
\begin{equation}
    ds^2 = g_{MN} dx^M dx^N = e^{-2T(y)}\eta_{\mu\nu} dx^\mu dx^\nu - dy^2\;,
\end{equation}
where $\eta_{\mu\nu} = {\rm diag}(1,-1,-1,-1)$. Under this ansatz, the Ricci scalar is $R = 20T'^2 - 8 T''$ and $\sqrt{g_i} = \left.\sqrt{g}\right|_{y_i}$.
Here the prime indicates the derivative respective to $y$. 

Since we are looking for the 4D Lorentz invariant background field solutions, let us assume that they only depend on 
the coordinate $y$.
The bulk equations of motion for the system are then
\begin{subequations}
    \begin{align}
       & 4 T'^2 - T''  = - \frac{2 \kappa^2}{3} V\;, \label{eq:t1} \\
       & T'^2  =  \frac{1}{12}  a'^2 - \frac{\kappa^2}{6} V \;,\label{eq:t2}\\
       & a''  =  4T' a' + \kappa^2 \frac{\partial V}{\partial a} \label{eq:a} \;.
    \end{align} 
\end{subequations}
The first two lines are obtained from the Einstein equation and the last one is the Klein-Gordon equation.
The bulk potential for the axion $a$ is given by 
\begin{equation}
    V = \Lambda_5  - \epsilon \frac{2 k^2}{\kappa^2} a^2\;, \quad \Lambda_5 = - \frac{6k^2}{\kappa^2}\;.
    \label{eq:v_b}
\end{equation}
Here, the bulk cosmological constant $\Lambda_5$ is set to a negative value so that we would have a pure AdS bulk solution if we neglected $a$.
In addition, $V$ contains the axion mass term, where $k$ is the AdS curvature scale with mass dimension one, and the $\epsilon$ suppression comes from the explicit breaking of the global symmetry that gives rise to $a$. 
This tachyonic mass parameter can originate from the full bulk potential of the axion, explained in appendix~\ref{appendix:bulk_potential}.
The numerical factors in the mass term are chosen for later convenience.
The boundary conditions are
\begin{subequations}
    \begin{align}
        2\left.T'\right|_{y_0,y_1} & = \pm \left.\frac{\kappa^2}{3} V_{0,1}(a) \right|_{y_0,y_1} \;, \label{eq:bc_t} \\ 
        2 \left.a'\right|_{y_0,y_1} & = \pm\left. \kappa^2 \frac{\partial V_{0,1}}{\partial a}\right|_{y_0,y_1}\;, \label{eq:bc_a}
    \end{align}
\end{subequations}
where the $+/-$ sign is for the UV/IR brane.
The boundary potentials of $a$ can be generally written as
\begin{equation}
    V_i(a) = \Lambda_i + \epsilon_i \frac{k}{\kappa^2}\left[ 1 - \cos(a - v_i)\right]\;,
    \label{eq:bp_a}
\end{equation}
with the bare brane tension $\Lambda_i$.
The potential $V_i(a)$ is periodic in $a$ and the $a$-dependent part is suppressed by a dimensionless parameter $\epsilon_i$.
In general, one has $\epsilon_1\neq \epsilon_2$ and $v_0\neq v_1$ because the potentials on two different brands arise from operators like $g_i \phi^N$ and $g_1 \neq g_2$.
Including other effects like potentially coupled strong dynamics or quantum gravity, one may have very different periodic brane potentials for $a$. 

The 4D effective potential for the radion/dilaton is obtained by integrating the action~\eqref{eq:action} along the $y$-direction, which gives
\begin{equation}
    V_{\rm eff} = \left.\left(\frac{6}{\kappa^2} \sqrt{g} T' \right) \right|_{y_0}^{y_1} + \sqrt{g_0} V_0|_{y_0} +  \sqrt{g_1} V_1|_{y_1}\;,
\end{equation}
where the bulk equation of motion eq.~\eqref{eq:t1} and eq.~\eqref{eq:t2} are applied.
Hence, the radion potential can be written collectively as
\begin{align}
    &V_{\rm eff} = V_{\rm UV} + V_{\rm IR}\;, \label{eq:V_eff_formal} \\[1ex]
    &V_{\rm UV}  = e^{-4T(y_0)} \left[ V_0(a(y_0)) - \frac{6}{\kappa^2} T'(y_0) \right] \;,\nonumber \\[1ex] 
    &V_{\rm IR}  = e^{-4T(y_1)} \left[ V_1(a(y_1)) + \frac{6}{\kappa^2} T'(y_1) \right] \;. \nonumber
\end{align}
Note that imposing the boundary condition for $T'$~\eqref{eq:bc_t} leads to the potential value where the radion sits at its vacuum solution and the radion does not contribute to the 4D cosmological constant.
Hence, to obtain the full shape of the potential, we do not impose eq.~\eqref{eq:bc_t}.
As the effective potential~\eqref{eq:V_eff_formal} is rather formal, to have a clear physical intuition, we need explicit solutions.

\subsection{The CPR solution}

The bulk equations of motion~\eqref{eq:t1}, \eqref{eq:t2}, \eqref{eq:a} can be analyzed analytically in the limit of the small bulk axion mass~\eqref{eq:v_b}, which is the solution suggested by Contino, Pomarol and Rattazzi (CPR).
We first review the CPR solution following ref.~\cite{Bellazzini:2013fga}, where hard (Dirichlet) boundary conditions for the bulk scalar are adopted.
For simplicity, we define the dimensionless coordinate $z = e^{-k y}$, under which the location of the UV brane is $\mu_0 = e^{-k y_0}$ and that of the IR brane is $\chi = e^{-k y_1}$.
Then, stabilizing the extra dimension is equivalent to giving $\chi$ a stable vacuum solution for a fixed $\mu_0$. 
We take $\mu_0=1$ ($y_0=0$) to simplify the following discussion.
Since $y_1\geq 0$, the dimensionless radion field $\chi$ takes $ 0 < \chi \leq  1$.

Let us first consider the massless limit of $\epsilon \to 0$. 
The bulk equations of motion for this case can be solved analytically, which are~\cite{Csaki:2000wz}
\begin{subequations}
    \begin{align}
        T(z) & = - \frac{1}{4}\log \left[ z^4 \left( \frac{1-\delta^8 z^{-8}}{1-\delta^8} \right) \right] \;, \label{eq:t_massless}\\ 
        a(z) & = \tilde{a} - \frac{\sqrt{3}}{2}\log \frac{z^4 - \delta^4}{z^4 + \delta^4}\;, \label{eq:a_massless}
    \end{align}
\end{subequations}
where $\tilde{a}$ and $\delta$ are two integration constants, we have taken the normalization, $T(z=1) =0$, and $\delta$ labels the location of a singularity. We then put the IR brane before the singularity to avoid the complication, $\chi \gtrsim \delta$.
This massless limit tells us the following:
\begin{enumerate}[label=\arabic*),leftmargin=15pt]
    \item 
    For the region close to the UV brane, $1\gtrsim z\gg \chi$,
    we have $a(z) \sim a_0 $ and $T \approx -\log z$, where $a_0$ is some UV boundary value.
    This is approximately a pure AdS solution.
    \item For the region close to the IR brane, $1 \gg z \gtrsim \chi$, $4T'a'$ dominates in eq.~\eqref{eq:a}
    and the solution significantly deviates from the pure AdS, which is reasonable because this region is close to the singularity $\delta$.
\end{enumerate}
Therefore, the bulk can be decomposed into two regions.
One is close to the UV brane where the solution remains close to the pure AdS,
and the other is close to the IR brane dominated by condensate.
We then call them `running region' and `condensate region' respectively.
After obtaining solutions in these two regions separately, we match them to get the full solution.

First, consider the running region. The solution is close to the pure AdS,
which means that $T\approx -\log z$ and $a(z)$ varies slowly, so that one can neglect the second-order derivative term in eq.\eqref{eq:a}.
The equation of motion for $a$ is expressed in terms of the coordinate $z$ as
\begin{equation}
    z \frac{da}{dz} + \epsilon a = 0\;.
\end{equation}
Then, the running region solutions (labeled by subscript `r') are given by
\begin{subequations}
    \begin{align}
        T_{\rm r} (z) & = - \log z \;,\\
        a_{\rm r} (z) & = a_0 z^{-\epsilon} \;.
    \end{align}
\end{subequations}
In the condensate region, the mass term in the bulk potential suppressed by a small $\epsilon$ can be neglected, and the condensate region solutions (labeled by subscript `c') are the same as that of the massless limit given in eqs.~\eqref{eq:t_massless} and~\eqref{eq:a_massless},
\begin{subequations}
    \begin{align}
        T_{\rm c}'(z)&= -k z \frac{dT_{\rm c}}{dz} = k \frac{z^8 + \delta^8}{z^8 - \delta^8} \;, \\
        a_{\rm c}(z) & = a_{\rm m} -\frac{\sqrt{3}}{2}\log \frac{z^4- \delta^4}{z^4 + \delta^4} \;,\label{eq:a_c}
    \end{align}
\end{subequations}
where $a_{\rm m}$ is a constant to be matched.
Note that $T_{\rm c}'(z)$ is the derivative of $T_{\rm c}$ with respect to $y$ but is expressed in $z$.
The matching condition is~\cite{Bellazzini:2013fga}
\begin{equation}
    \lim_{z\to \infty} a_{\rm c} = \lim_{z \to \chi} a_{\rm r} \implies a_{\rm m} = a_0 \chi^{-\epsilon}\;.
\end{equation}
Since $T_{\rm r} = \lim_{z\gg \delta}T_{\rm c}$, the matching for $T_{\rm c}$ and $T_{\rm r}$ is trivial.
Suppose the boundary values are
$a(1) \approx a_{\rm r}(1) = \tilde v_0$ and
$a(\chi) \approx a_{\rm c}(\chi) = \tilde v_1$, then
\begin{subequations}
    \begin{align}
        a_0  & = \tilde v_0 \;,\\
        \delta^4  & = \chi^4 \tanh \beta\;,
    \end{align} \label{eq:actual_values}
\end{subequations}
where the function $\beta$ is 
\begin{equation}
    \beta(\chi) = \frac{1}{\sqrt{3}} \left( \tilde v_1 - \tilde v_0  \chi^{-\epsilon} \right)\;.
    \label{eq:beta}
\end{equation}
The full approximate solution is $a(z) \approx a_{\rm r}(z) + a_{\rm c}(z) - a_{\rm m}$. 
Explicitly, we have
\begin{subequations}
    \begin{align}
        a(z) & \approx \tilde v_0 z^{-\epsilon} - \frac{\sqrt{3}}{2} \log \left(-1 + \frac{2z^4}{z^4 + \chi^4 \tanh \beta} \right)\;, \\ 
        T'(z) & \approx k \left( - 1 + \frac{2z^8}{z^8 - \chi^8 \tanh^2 \beta} \right)\;.\label{eq:tp_full}
    \end{align}
\end{subequations}
Therefore, the UV and IR parts of the effective potential for the radion are 
obtained by putting eq.~\eqref{eq:tp_full} back to eq.~\eqref{eq:V_eff_formal}, which is
\begin{subequations}
    \begin{align}
        V_{\rm UV} & = \mu_0^4 \left[ V_0(\tilde v_0) + \frac{\Lambda_5}{k} \right] \;, \\ 
        V_{\rm IR} & = \chi^4  \left[ V_1(\tilde v_1) - \frac{\Lambda_5}{k} \cosh(2\beta) \right] {\rm sech}^2 \beta \;,
    \end{align}
\end{subequations}
where $V_i(\tilde v_i)$ gives the effective brane tension including the bare tension $\Lambda_i$
and a potential energy from the axion $a$.
Several comments are in order. 
For the Dirichlet boundary condition with $\tilde v_i \to v_i$, we have $V_1(\tilde v_1) \to \Lambda_1$, which reduces to exactly the CPR scenario.
For a general $\mu_0 \neq 1$, one can replace $\chi \to \chi/\mu_0$ in $\beta$~\eqref{eq:beta} to obtain the $V_{\rm eff}$.
Note that $V_{\rm UV}$ does not contain $\chi$, so it is a constant contribution and does not determine the stabilization of the radion. We hence focus on $V_{\rm IR}$.

\subsection{Soft boundary conditions}

Since the boundary potential of the axion field $a$ is suppressed by a small parameter $\epsilon_i$ as in eq.~\eqref{eq:bp_a}, there is a potential energy shift due to the mismatch between the actual field value $\tilde v_i$ and the boundary parameter $v_i$,
where $\left.\partial V_i/\partial a \right|_{a= v_i} = 0 $.

The boundary conditions~\eqref{eq:bc_a} with the boundary potentials~\eqref{eq:bp_a} tell us that
\begin{subequations}
    \begin{align}
        2\epsilon \tilde v_0 & = \epsilon_0  \sin \left( \tilde v_0 - v_0 \right) \;, \label{eq:v_0'} \\ 
        - 4\sqrt{3}  \sinh(2\beta) & = \epsilon_1  \sin\left( \tilde v_1 - v_1 \right) \;, \label{eq:v_1'}
    \end{align}
\end{subequations}
where we have used eq.~\eqref{eq:actual_values}.
The first equation~\eqref{eq:v_0'} indicates that $\tilde v_0$ is a function of $\{\epsilon, \epsilon_0, v_0\}$, and $|\tilde v_0| \leq |\epsilon_0/2\epsilon|$. 
Given a value of $\tilde v_0$, one can always find at least one set of $\{\epsilon_0, v_0\}$ that satisfies eq.~\eqref{eq:v_0'} for a fixed $\epsilon$, and $\beta$ depends on $\{\epsilon_0, v_0\}$ only through $\tilde v_0$.
Hence, from now on we take $\tilde v_0$ as a free parameter instead of using $\{\epsilon_0, v_0\}$.
The second equation~\eqref{eq:v_1'} gives the expression for the IR boundary potential value, which is
\begin{equation}
    V_1(\tilde v_1) = \Lambda_1 + \epsilon_1 \frac{k}{\kappa^2} \left[ 1- \eta \sqrt{1- \frac{48}{\epsilon_1^2} \sinh^2 (2\beta)} \right]\;,
\end{equation}
with $\eta$ being a sign factor which depends on the value of $\alpha \equiv \tilde v_1 -v_1 \mod{2\pi}$.
Explicitly, $\eta = 1$ for $0 \leq \alpha < \pi/2$ or $ 3\pi/2 \leq \alpha < 2\pi$, while $\eta = -1$ for $\pi/2 \leq \alpha < 3\pi/2$.
The effective potential for the radion in the unit of $|\Lambda_1|$ is then given by
\begin{align}
    &V_{\rm eff}(\chi)  = \chi^4 F[\beta(\chi)] \label{eq:V_eff} \;, \\[1ex]
    &F[\beta(\chi)]  = \left[{\rm sgn}(\Lambda_1) + \xi \Delta(\beta) + \xi \cosh(2\beta) \right] {\rm sech}^2 \beta \;, \nonumber \\
    &\Delta(\beta)  = \frac{\epsilon_1}{6}\left[ 1- \eta \sqrt{1- \frac{48}{\epsilon_1^2} \sinh^2 (2\beta)} \right]\;. \nonumber
\end{align}
Here, the dimensionless parameter $\xi = -\Lambda_5 / k |\Lambda_1|>0$ describes the relative strength between the bulk cosmological constant and the IR brane tension.
The value $\tilde v_1$ can be solved from eq.~\eqref{eq:v_1'} in the limit of a small $\epsilon_1$ together with eq.~\eqref{eq:beta},
which gives the explicit form of the function $\beta$,
\begin{equation}
    \beta(\chi) = \frac{\epsilon_1}{8\sqrt{3}} \sin \left( v_1 - \tilde v_0 \chi^{-\epsilon} \right) + \mathcal{O}\left( \epsilon_1^2 \right) \;.    \label{eq:beta_epsilon_1}
\end{equation}
The effective potential \eqref{eq:V_eff} will stabilize the radion when the parameters $\{\epsilon,\epsilon_1, v_1, \tilde v_0,\xi\}$ are chosen properly.

\section{Multiple vacua}
\label{sec:VEVs}

The VEV of the radion $\langle \chi \rangle$ is determined by the equation $dV_{\rm eff}/ d\chi =0$,
which is explicitly written as
\begin{equation}
    \langle\chi\rangle^3 \left[ 4 F(\langle\chi\rangle) + \langle\chi\rangle \left.\frac{d F}{d \chi}\right|_{\langle\chi\rangle}\right] = 0 \;.
    \label{eq:veveq}
\end{equation}
The VEV is determined by the parameter $\xi$ and the form of the function $\beta$.
The radion mass can be obtained by expanding the effective potential around the VEV,
\begin{equation}
    m_\chi^2 = \left.\frac{d^2 V_{\rm eff}}{d \chi^2} \right|_{\langle\chi\rangle} = \langle\chi\rangle^2 \left[ \langle\chi\rangle^2 \left.\frac{d^2F}{d\chi^2}\right|_{\langle\chi\rangle} - 20 F(\langle\chi\rangle) \right]\;,
\end{equation}
in the unit of $|\Lambda_1|^{1/4}$.
The physical mass receives a normalization factor from the kinetic term.
Parameterizing a perturbation of the IR brane tension as $\xi = 1+\sigma$, we will see that the behavior of the effective radion potential significantly depends on the sign of $\sigma$.
We start with the case of $\sigma =0$ and then discuss the case of $\sigma \neq 0$.

\subsection{Tuned IR brane tension}

\begin{figure}
    \centering
    \includegraphics[width=7cm]{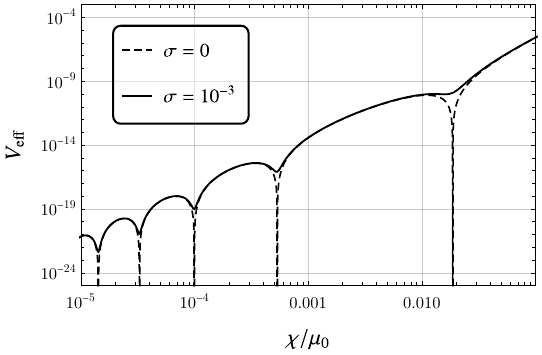}
    \caption{The radion effective potential (eq.~\eqref{eq:V_eff} and eq.~\eqref{eq:beta_epsilon_1}) with a perturbation of the IR brane tension, $\xi=1 + \sigma$. Here we choose $\epsilon = 0.3$,  $\epsilon_1=0.2$, $\tilde v_0 = 1$ and $v_1 = 3.3$. The dashed line corresponds to the case of $\sigma=0$, whose energy at each local minimum extends to zero. The solid line represents the case of a positive mis-tuning, $\sigma = 10^{-3}$, which gives a positive energy at each local minimum. }
    \label{fig:V_eff_sigma_0p}
\end{figure}

The tuning between the bare IR brane tension and the bulk cosmological constant, $\Lambda_1 = \Lambda_5/k$, indicates that $\xi=1$ and ${\rm sgn}(\Lambda_1) = -1$.
Denoting $F_0 \equiv F[\xi =1, \Lambda_1<0]$, the VEV equation~\eqref{eq:veveq} becomes
\begin{equation}
    G_0[\beta,\chi] = 4F_0[\beta] + \chi \beta'(\chi) \frac{dF_0}{d\beta} = 0\;.
    \label{eq:veveq_0}
\end{equation}
Since $F_0[\beta = 0]  \propto (1-\eta)$  and $ \left.dF_0/d\beta \right|_{\beta=0} = \Delta'(0) = 0$, it is straightforward to see that $\beta=0$, $\eta=1$ is a solution to eq.~\eqref{eq:veveq_0}.
The stability of the solution is determined by the sign of the radion mass $m_\chi^2$.
For $\beta=0$, it is
\begin{align}
    &m_\chi^2(\beta\to 0)  = \frac{10}{3}\epsilon_1 (\eta -1 ) {\langle\chi\rangle}^2 \label{eq:m_chi_square} \\
    & \quad + \frac{{\langle\chi\rangle}^4 \beta'({\langle\chi\rangle})^2}{3\epsilon_1} \left[ 96 \eta + 12 \epsilon_1 + \epsilon_1^2 (\eta-1) \right]\nonumber \;.
\end{align}
Then, $\eta = 1$ gives $m_\chi^2 >0$.
Since the $\beta(\chi)$ is formed by a sinusoidal function, which is periodic,
we have multiple VEV solutions that correspond to $\beta(\langle \chi \rangle^{(p)})=0$ 
and $\eta(\langle \chi \rangle^{(p)}) = 1$, 
which are labeled by an integer $p\in \mathbb{Z}$,
\begin{equation}
     \langle \chi \rangle^{(p)} =  \left( \frac{\tilde v_0}{v_1 - 2p \pi} \right)^{1/\epsilon}   \;.
\end{equation}
To have real-valued solutions for the physical VEVs, the denominator and numerator in the parentheses should take the same sign, which indicates that $\tilde v_0 (v_1 - 2p \pi) >0$.
Together with the requirement $\langle \chi \rangle < 1$, the integer $p$ is constrained.
For $\tilde v_0>(<)0$, we have $2p \pi<(>)v_1- \tilde v_0$, and a larger $p$ corresponds to a larger (smaller) VEV.
In both cases, the number of vacuum solutions is infinite.
Defining the critical $p_{\rm crit} = {\rm Int}[(v_1 - \tilde v_0)/2\pi]$, where ${\rm Int}[X]$ is the integer part of a real number $X$, $\langle \chi \rangle^{(p_{\rm crit})}$ is the largest VEV.
The radion mass-to-VEV ratio around the $p$th VEV is, in the small $\epsilon_1$ limit,
\begin{equation}
    \left[\frac{m_\chi}{\langle \chi \rangle}\right]^{(p)} = \frac{\epsilon \epsilon_1^{1/2}}{\sqrt{6}}\left|v_1 - 2p \pi \right| + \mathcal{O}(\epsilon \epsilon_1^{3/2})\;.
\end{equation}
The potential values of all VEVs are vanishing because 
\begin{equation}
    \langle V_{\rm eff} \rangle^{(p)} \propto F_0[\beta = 0, \eta=1] = 0\;.
\end{equation}
The vacua are degenerate as indicated by the dashed line in Fig.~\ref{fig:V_eff_sigma_0p}.

\subsection{Perturbed IR brane tension}

\begin{figure}
    \centering
    \includegraphics[width=7cm]{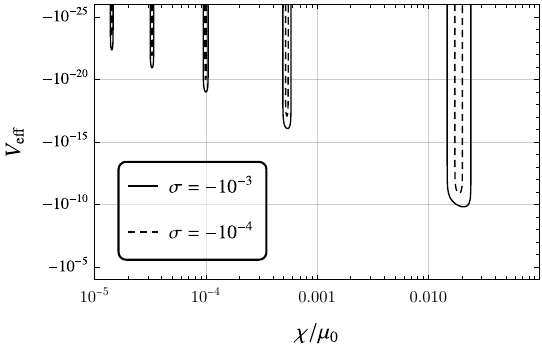}
    \caption{The radion effective potential (eq.~\eqref{eq:V_eff} and eq.~\eqref{eq:beta_epsilon_1}) with a perturbation of the IR brane tension, $\xi=1 + \sigma$ with $\sigma<0$. Here we choose $\epsilon = 0.3$,  $\epsilon_1=0.2$, $\tilde v_0 = 1$ and $v_1 = 3.3$.
    The solid line stands for $\sigma = - 10^{-3}$ and the dashed line is for $\sigma = -10^{-4}$.}
    \label{fig:V_eff_sigma_n}
\end{figure}

We now consider a perturbation of the IR brane tension, $\xi= 1 + \sigma$ with $|\sigma|\ll 1$
(still focusing on ${\rm sgn}(\Lambda_1)=-1$).
The effective quartic coupling $F[\beta(\chi)]$ in eq.~\eqref{eq:V_eff} can be separated as
\begin{equation}
    F[\beta(\chi)]= F_0[\beta] + \sigma \left( F_0[\beta] + {\rm sech}^2 \beta \right)\;.
    \label{eq:F_sigma}
\end{equation}
The VEV equation~\eqref{eq:veveq} then becomes
\begin{align}
    &G_0[\beta,\chi] =  \label{eq:veveq_sigma}\\
    &- \sigma \left( G_0[\beta,\chi] + 4 {\rm sech}^2\beta - 2 \chi \beta'(\chi) {\rm sech}^2 \beta \tanh \beta \right) \;.\nonumber
\end{align}
This equation cannot be solved exactly. In the limit of a small $\sigma$, the perturbation method gives
\begin{equation}
    \langle \chi \rangle = x_0 - \sigma \frac{4}{x_0\left[\beta'(x_0)\right]^2\left(4+ 32/\epsilon_1 \right)} + \mathcal{O}(\sigma^2)\;,
\end{equation}
where $x_0$ satisfies $\beta(x_0) = 0$ from the $\sigma^0$ order equation, $G_0[\beta(x_0),x_0]=0$.
Since there are multiple $x_0^{(p)}$ satisfying the equation, so does $\langle \chi \rangle^{(p)}$.
Taking the leading order of $\epsilon_1$ in $\beta$~\eqref{eq:beta_epsilon_1}, we obtain
\begin{equation}
    \langle \chi \rangle^{(p)} \approx \left( \frac{\tilde v_0}{v_1 - 2p \pi} \right)^{1/\epsilon} \left[ 1 - \frac{ 24 \sigma}{\epsilon^2 \epsilon_1 (v_1 - 2p \pi)^2} + \mathcal{O}(\sigma^2) \right]\;.
    \label{eq:vev_sigma}
\end{equation}
For the expansion series in $\sigma$ to be valid, one needs $|\sigma| \ll \epsilon^2 \epsilon_1/24$ given $v_1- 2p \pi = \mathcal{O}(1)$, which is extremely small for small bulk $\epsilon$ and IR boundary $\epsilon_1$.
The mass-to-VEV ratio is given by, taking the leading order of $\epsilon_1$ in $\beta$~\eqref{eq:beta_epsilon_1},
\begin{equation}
    \left[\frac{m_\chi}{\langle \chi \rangle}\right]^{(p)} \approx \frac{\epsilon \epsilon_1^{1/2}}{\sqrt{6}}\left|v_1 - 2p \pi \right| + \mathcal{O}(\sigma)\;.
    \label{eq:mass-to-VEV_sigma}
\end{equation}
The locations of the VEVs are slightly shifted by $\sigma$.
The potential energy at each minimum is 
\begin{equation}
    \langle V_{\rm eff} \rangle^{(p)} \approx \sigma \left( \frac{\tilde v_0}{v_1 - 2p  \pi}\right)^{4/\epsilon} + \mathcal{O}(\sigma^2)\;.
\end{equation}
Note that now the potential energy depends on $\sigma$, which indicates two distinct scenarios
with different potential applications:
\begin{enumerate}[label=\arabic*),leftmargin=15pt]
    \item For $\sigma>0$, the potential energy at every local minimum is positive, $\langle V_{\rm eff}\rangle^{(p)}>0$.
    A larger VEV then corresponds to a larger vacuum energy as indicated by the solid line in Fig.~\ref{fig:V_eff_sigma_0p}.
    This dilaton/radion potential, together with a constant contribution from $V_{\rm UV}$,
    could be used for the relaxion scenario~\cite{Graham:2015cka}.
    Suppose the Standard Model electroweak sector lives on the IR brane.
    The electroweak scale is warped down and essentially given by a dilaton/radion VEV.
    Each VEV corresponds to an electroweak scale $M_{\rm EW}^{(p)} \sim M_* \langle \chi \rangle^{(p)} $,
    where $M_*$ is the fundamental scale in the 5D theory, which is close to the 4D Planck scale.
    Consider $\tilde v_0>0$ for simplicity. 
    One may start with compactification with a large radion VEV (a large $p$) in the early Universe.
    This is a false vacuum because a smaller VEV with a smaller energy exists.
    Through a phase transition, the Universe evolves to a smaller VEV, achieving a lower electroweak scale.
    There are an infinite number of vacua with smaller energies.
    One then needs a constant UV piece $V_{\rm UV}$ in eq.~\eqref{eq:V_eff_formal}
    to terminate the cascade process dynamically.~\footnote{The nucleation in the AdS is terminated~\cite{Hebecker:2020aqr}.}
    With a proper choice of parameters, a relaxion model that explains the electroweak scale is expected to be constructed. 
    A detailed exploration of this possibility is left for a future study.
    
    \item For $\sigma<0$, the potential energy at every local minimum is negative, $\langle V_{\rm eff}\rangle^{(p)}<0$,
    and a larger radion VEV corresponds to a smaller energy, as shown in Fig.~\ref{fig:V_eff_sigma_n}.
    This indicates that the potential $V_{\rm eff}(\chi)$ has a well-defined nonzero global minimum, which is the largest VEV at $p_{\rm crit}$. 
    The radion will be eventually stabilized at $\langle \chi \rangle^{(p_{\rm crit})}$.
    Note that in this vacuum, the radion mass is suppressed as $m_\chi \propto \epsilon \sqrt{\epsilon_1}$~\eqref{eq:mass-to-VEV_sigma}.
    To estimate the lower limit of the suppression, we consider the largest VEV which roughly gives $|v_1 - 2p_{\rm crit} \pi| = \mathcal{O}(1)$. 
    We know that the perturbation series~\eqref{eq:vev_sigma} is valid until $\epsilon^2 \epsilon_1 \sim 24 \sigma$.
    Hence we have
    \begin{equation}
        \frac{m_\chi}{\langle \chi \rangle} \sim \sqrt{\frac{\epsilon^2 \epsilon_1}{6}} > 2\sqrt{|\sigma|} \;.
    \end{equation}
    Note that going beyond this limit requires a detailed numerical evaluation.
\end{enumerate}

Here our radion mass scales as $m_\chi/\langle \chi \rangle \propto \epsilon$, which is the same as in ref.~\cite{Goldberger:1999un}, in the perturbative region, where $|\sigma|\ll \epsilon^2 \epsilon_1/24$. This linear dependence in $\epsilon$ results from the tuning between the bulk cosmological constant and the IR brane tension, $\xi \to 1$. There would be extra terms that $\propto \beta'(\langle \chi \rangle)$ in $m_\chi^2$~\eqref{eq:m_chi_square}, and VEV equation would give $\beta(\langle \chi \rangle) \neq 0$ for $\xi\neq 1$. 
For large mis-tuning $\sigma$, one should do the perturbative analysis in small $\epsilon$ instead of $\sigma$, which would give the leading scaling as $m_\chi/\langle \chi \rangle \propto \epsilon^{1/2}$ as in the CPR scenario~\cite{Bellazzini:2013fga}.

\subsection{Beyond perturbation}

Naively from Fig.~\ref{fig:V_eff_sigma_0p}, one can see that for a larger positive $\sigma$,
the potential goes upwards and the wiggly behavior persists with a smaller amplitude.
For negative $\sigma<0$ with a sufficiently small $|\sigma|$, the potential is still bounded from below
(see Fig.~\ref{fig:V_eff_sigma_n}) and has a global minimum as analyzed in the last subsection.
However, for $|\sigma| \geq \epsilon^2 \epsilon_1/24$, one cannot perturbatively solve the VEV equation~\eqref{eq:veveq}.
In this subsection, we give a numerical analysis that goes beyond the small $\sigma$ perturbation. 

\begin{figure}
    \centering
    \includegraphics[width=7cm]{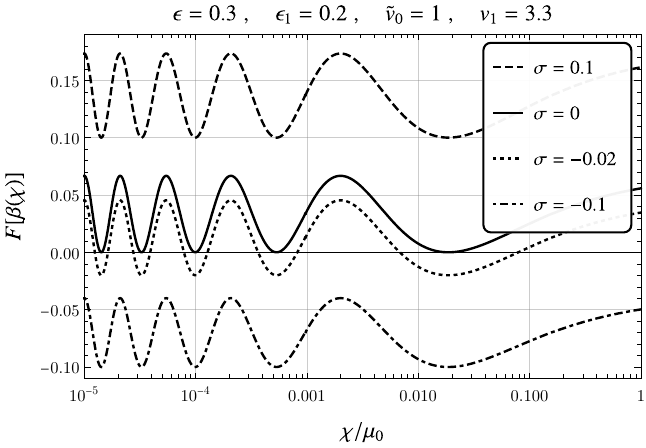}
    \caption{The coupling $F[\beta(\chi)]$ in eq.~\eqref{eq:V_eff} with the $\beta$ of eq.~\eqref{eq:beta_epsilon_1}
    in the small $\epsilon_1$ limit. Here we choose $\epsilon=0.3$, $\epsilon_1=0.2$, $\tilde v_0 = 1$ and $v_1 =3.3$.}
    \label{fig:beta_sigma}
\end{figure}

The behavior of the radion potential is determined by the function $F[\beta(\chi)]$ as shown in eq.~\eqref{eq:V_eff}.
The following analysis adopts the expression of $\beta$ in the small limit of $\epsilon_1$
given in eq.~\eqref{eq:beta_epsilon_1}. 
From the analytical expression, one can already see that the coupling $F$ fluctuates with $\chi$ since $\beta$ is bounded by a sinusoidal function.
As shown in Fig.~\ref{fig:beta_sigma}, for the case of $\sigma=0$, the coupling $F$ is always non-negative and hits the zero at $\langle \chi \rangle^{(p)}$, so that the radion potential is bounded from below.
For $\sigma>0$, the coupling $F$ is always positive, and the lower bound $F \geq \sigma$ from eq.~\eqref{eq:F_sigma} indicates that the effective potential $V_{\rm eff} = \chi^4 F$ is bounded.
All local minima $\langle \chi \rangle^{(p)}$ are metastable since the only global minimum is $\chi \to 0$. 
For a slightly negative mis-tune, for example, $\sigma= - 0.02$ in Fig.~\ref{fig:beta_sigma}, the coupling $F$ fluctuates around zero. In this case, as long as $F(\chi = 1)>0$ is satisfied, the potential is bounded from below and the critical VEV $\langle \chi \rangle^{(p_{\rm crit})}$ is the global minimum of the potential.
The situation becomes subtle when the mis-tune goes large in the negative direction, which makes the coupling negative, $F<0$ (see the $\sigma = -0.1$ curve in Fig.~\ref{fig:beta_sigma}). 
Since the radion potential is only defined at $0< \chi \leq 1$, as long as $V_{\rm eff}(\chi =1)$ is not the global minimum, the potential is bounded from below, and the $\langle \chi \rangle^{(p_{\rm crit})}$ is still the globally stable VEV solution.
Analytically, the $F$ takes extrema at $\beta=0$ where those with $\eta=1$ correspond to minima and those with $\eta=-1$ give maxima. 
From eq.~\eqref{eq:F_sigma}, one can see that
\begin{equation}
    F_{\rm min} = \sigma\;,\quad F_{\rm max} = \frac{\epsilon_1}{3} + \sigma \left( \frac{\epsilon_1}{3} + 1 \right)\;.
\end{equation}
The VEV can be approximated by the location of $F_{\rm min}$ for a small $|\sigma|$
as given in eq.~\eqref{eq:vev_sigma}.
For positive (negative) $\sigma$, the actual VEVs take values $\langle \chi \rangle^{(p)}$ smaller (larger) than the location of $F_{\rm min}$, indicated by the general form of the potential.

\begin{figure}
    \centering
    \includegraphics[width=7cm]{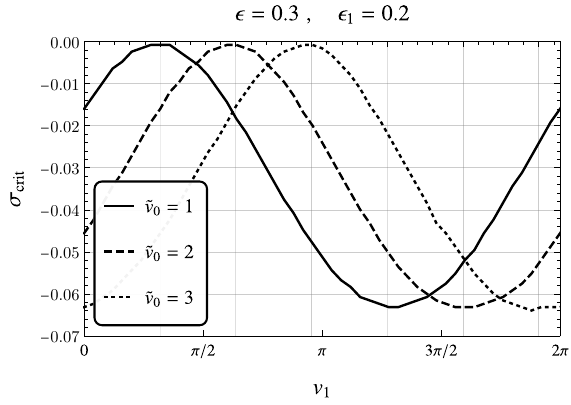}
    \caption{The critical point $\sigma_{\rm crit}$, describing the stability of the effective potential \eqref{eq:V_eff},
    as a function of $v_1$ with different $\tilde v_0$. Here we choose $\epsilon =0.3$ and $\epsilon_1=0.2$, and use the $\beta$ of eq.~\eqref{eq:beta_epsilon_1} in the small $\epsilon_1$ limit.}
    \label{fig:sigma_v1_diff_v0}
\end{figure}

\begin{figure}
    \centering
    \includegraphics[width=7cm]{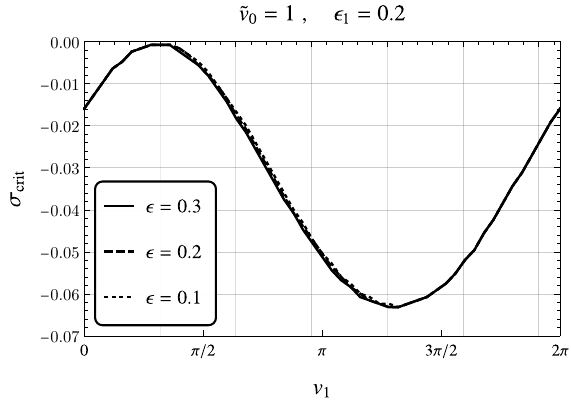}
    \caption{The critical point $\sigma_{\rm crit}$, describing the stability of the effective potential \eqref{eq:V_eff}, 
    as a function of $v_1$ with different $\epsilon$.
    Here we choose $\tilde v_0=1 $ and $\epsilon_1=0.2$,
    and use the $\beta$ of eq.~\eqref{eq:beta_epsilon_1} in the small $\epsilon_1$ limit.}
    \label{fig:sigma_v1_diff_epsilon}
\end{figure}

\begin{figure}
    \centering
    \includegraphics[width=7cm]{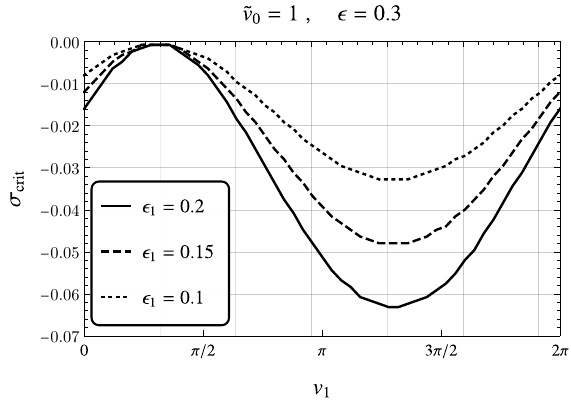}
    \caption{The critical point $\sigma_{\rm crit}$, describing the stability of the effective potential \eqref{eq:V_eff},
    as a function of $v_1$ with different $\epsilon_1$.
    Here we choose $\tilde v_0=1 $ and $\epsilon=0.3$,
    and use the $\beta$ of eq.~\eqref{eq:beta_epsilon_1} in the small $\epsilon_1$ limit.}
    \label{fig:sigma_v1_diff_epsilon1}
\end{figure}

For fixed $\tilde v_0$, $v_1$, $\epsilon$ and $\epsilon_1$, there exists a critical point $\sigma_{\rm crit}$ such that for all $\sigma> \sigma_{\rm crit}$, one has $V_{\rm eff}(\chi\to 1) > \langle V_{\rm eff} \rangle^{(p_{\rm crit})}$, which means that the VEV solution is globally stable.
For $\sigma< \sigma_{\rm crit}$, the potential is unstable, which indicates the collapse of the extra dimension.
Hence, the point $\sigma_{\rm crit}$ determines the stability of the radion potential.
Using the formula~\eqref{eq:V_eff} and the expression of $\beta$ in eq.~\eqref{eq:beta_epsilon_1} in the limit of a small $\epsilon_1$, we numerically find such $\sigma_{\rm crit}$.
Fig.~\ref{fig:sigma_v1_diff_v0} shows the relation between $\sigma_{\rm crit}$ and $v_1$ with different $\tilde v_0$, keeping $\epsilon=0.3$ and $\epsilon_1$ fixed.
The change of $\tilde v_0$ horizontally shifts the curve.
Note that the $v_1$ has the periodicity of $2\pi$ from eq.~\eqref{eq:beta_epsilon_1}.
The variation of the curve for the relation between $\sigma_{\rm crit}$ and $v_1$ with respect to different $\epsilon$ is shown in Fig.~\ref{fig:sigma_v1_diff_epsilon}, where we can see that the $\epsilon$ gives only a small effect. 
Fig.~\ref{fig:sigma_v1_diff_epsilon1} shows that the curve is scaled upwards for a smaller $\epsilon_1$.
This indicates that a smaller $\epsilon_1$ gives a larger $\sigma_{\rm crit}$.
Comparing Fig.~\ref{fig:sigma_v1_diff_v0}, Fig.~\ref{fig:sigma_v1_diff_epsilon} and Fig.~\ref{fig:sigma_v1_diff_epsilon1} with Fig.~\ref{fig:beta_sigma}, one can see that the $\sigma_{\rm crit}(v_1)$ always corresponds to the $F(\chi)$ curve that satisfies $F(\chi\to 1) >0$.
This indicates that $\sigma \ll -\epsilon_1/ (\epsilon_1+3)$ does not give a stable radion potential.

\begin{figure}
    \centering
    \includegraphics[width=7cm]{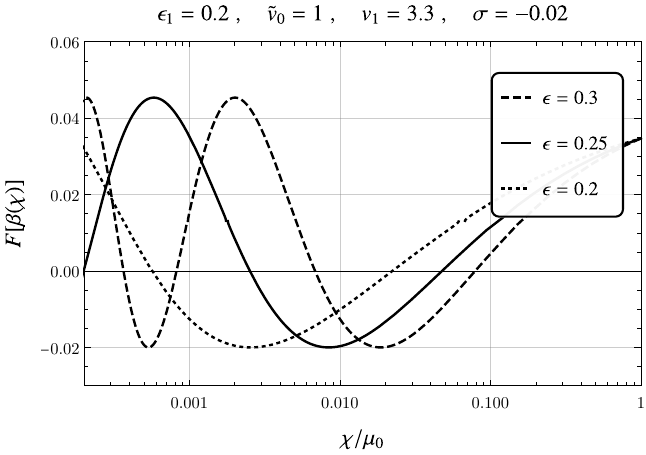}
    \caption{The coupling $F[\beta(\chi)]$ in eq.~\eqref{eq:V_eff} with the $\beta$ of eq.~\eqref{eq:beta_epsilon_1}
    in the small limit of $\epsilon_1$, as a function of $\chi$ for different $\epsilon$.
    Here we choose $\epsilon_1 = 0.2$, $\tilde v_0 = 1$ and $v_1 =3.3$.}
    \label{fig:beta_sigma_epsilon}
\end{figure}

Eq.~\eqref{eq:mass-to-VEV_sigma} indicates that the light dilaton is obtained by a small $\epsilon \epsilon_1^{1/2}$.
However, a smaller $\epsilon_1$ gives a smaller $|\sigma_{\rm crit}|$,
which shrinks the parameter space for $\sigma$ as shown in Fig.~\ref{fig:sigma_v1_diff_epsilon1}.
Moreover, a smaller $\epsilon$ will lead to a huge suppression on the value of $\langle \chi \rangle^{(p_{\rm crit})}$
as indicated by Fig.~\ref{fig:beta_sigma_epsilon},
where the location of the absolute minimum $\langle \chi \rangle^{(p_{\rm crit})}$
is pushed towards the left for a smaller $\epsilon$. 
This means that for a very small $\epsilon$, a huge fine-tuning between other parameters is required to have a reasonable $\langle \chi \rangle^{(p_{\rm crit})}$.
Therefore, an extremely light radion requires fine-tuning.


\section{Summary and discussion}
\label{sec:summary}

We have explored the dilaton/radion stabilization by a 5D axion-like field with realistic boundary potentials, which have
two general features: one is its smallness in terms of energy and
the other is the periodicity.
The smallness makes the boundary condition nearly Neumann
(deviating from the conventionally adopted Dirichlet condition),
and the periodicity leads to multiple locally stable vacuum solutions, giving us a wiggly dilaton potential.
The stability of such a wiggly dilaton potential against the mis-tuning of the IR brane tension has been discussed through a perturbation method and a numerical analysis. 
Here we employed an effective field theoretic computation to obtain the dilaton potential and mass. It will be interesting to see how it compares with the mass eigenvalues of the fluctuations described by the coupled dilaton-axion system using the techniques in ref.~\cite{Csaki:2000zn}.

Since the wiggly dilaton potential has a landscape of local minima,
which correspond to distinct scales for fields on the IR brane,
the dynamics of the dilaton field in cosmology may bring us interesting phenomena.
For example, a series of first-order phase transitions might happen if the dilaton was initially located in a false vacuum.
The nucleation dynamics will be very different since the scales inside and outside the bubble are hierarchical.

The stable domain wall exists if $\sigma=0$. At the transition region,
the scales of fields on the IR brane have position dependence.
This intriguing feature requires more understanding.

For a negative $\sigma< \sigma_{\rm crit}$, the dilaton potential is unstable, and the global minimum is the $\chi \to 1$. 
With a proper choice of parameters, one can make the dilaton with a flat enough potential which may serve as the inflaton. 
Meanwhile, the wiggly feature may be used to create large density fluctuations that populate primordial black holes.
Various applications of the wiggly dilaton deserve further studies. 

From an effective field theory perspective, the existence of a set of vanishing quartic couplings makes the dilaton potential wiggles~\cite{Coradeschi:2013gda}. 
From a 5D holographic picture, it is the periodicity of the IR boundary potential that generates the wiggly dilaton. However, it is difficult to identify its 4D origin.
It might originate from nontrivial VEVs at the IR of marginally-relevant operators that trigger SBSI, whose suppressed beta function is dual to the 5D bulk ALP profile. The explicit 4D construction is pending for future work.

\section*{Acknowledgments}

The work is supported by Natural Science Foundation of Shanghai. The work of Y. -C. Qiu is supported by the K. C. Wong Educational Foundation.

\appendix

\section{$\epsilon$ from a bulk axion potential}
\label{appendix:bulk_potential}

The tachyonic bulk axion mass term can originate from a bulk potential of the 5D axion field.
Here, we will show that the form of the obtained dilaton potential would be the same as eq.~\eqref{eq:V_eff}
if we took a full bulk axion potential instead of eq.~\eqref{eq:v_b}.

Let us consider the following complete bulk potential of the 5D axion field:
\begin{equation}
    V (a) = \Lambda_5 + \tilde \epsilon \frac{k^2}{\kappa^2} \left( 1- \cos a \right)\;,
\end{equation}
whose arbitrary phase is set to zero without loss of generality.
At the minimum, the potential value leads to the bulk cosmological constant $V(a\to 0) = \Lambda_5$.
Here $\tilde \epsilon>0$ is assumed to originate from an explicit breaking of a global symmetry.
Given this bulk potential, 
the bulk equation of motion for the $a(z)$~\eqref{eq:a} in terms of $z=e^{-ky}$ is 
\begin{equation}
    \frac{d^2 a}{d z^2} + \left( \frac{1}{z} - 4\frac{dT}{dz } \right) \frac{da}{d z} = \frac{\tilde \epsilon}{z^2}\sin a  \;.
\end{equation} 
For the running region, the metric function is given by $T_{\rm r} = - \log z$,
and the $a(z)$ evolves slowly without deviating too much from the initial value $a(\mu_0) = \tilde v_0$.
Then, the equation of motion in this region can be approximated as
\begin{equation}
    \frac{d^2 a}{d z^2} + \frac{5}{z}\frac{d a}{dz} = \frac{\tilde{\epsilon}}{z^2} \left[ \sin \tilde v_0 +  (a - \tilde v_0) \cos \tilde v_0  +\cdots \right]\;.
\end{equation}
We expand the potential term for a small deviation of $a$ from $\tilde v_0$ and keep the leading order.
The general solution is 
\begin{align}
    a_{\rm r}(z) &\approx \tilde v_0 - \tan \tilde v_0 \\
    & \quad + c_1 z^{-2+\sqrt{4 + \tilde \epsilon \cos \tilde v_0}} + c_2 z^{-2 - \sqrt{4+ \tilde \epsilon \cos \tilde v_0}}\;,\nonumber
\end{align}
where $c_{1,2}$ denote two integration constants.
In order to have a slow-evolving profile for $a_{\rm r}$, $d a_{\rm r}/ d z$ should almost vanish.
Then, we set $c_2 \approx 0$, because in the small $\tilde \epsilon$ limit,
the power $-2 -\sqrt{4 + \tilde \epsilon \cos \tilde v_0} \approx -4 - \tilde \epsilon \cos \tilde v_0/4 $ is not suppressed
while the other power $-2 +\sqrt{4 + \tilde \epsilon \cos \tilde v_0} \approx \tilde \epsilon \cos \tilde v_0/4 $ is parametrically small. 

The solution for the condensate region is the same as the case in the potential is neglected, given in eq.~\eqref{eq:a_c}.
The matching condition then indicates that
\begin{equation}
    a_{\rm m} = \tilde v_0 - \tan \tilde v_0 + c_1 \chi^{\epsilon}\;,
\end{equation}
where we have defined $\epsilon \equiv 2 - \sqrt{4+ \tilde \epsilon \cos \tilde v_0}$.
Similar to eq.~\eqref{eq:actual_values}, we have
\begin{subequations}
    \begin{align}
        c_1 & = \mu_0^{\epsilon} \tan \tilde v_0 \;, \\[1ex]
        \delta^4 & = \chi^4 \tanh \beta \;, \\
        \beta(\chi) &= \frac{1}{\sqrt{3}} \left[\tilde v_1 - \tilde v_0 + \tan \tilde v_0\left( 1+ \chi^{-\epsilon} \right) \right]\;.
    \end{align}
\end{subequations}
The boundary conditions~\eqref{eq:bc_a} give
\begin{subequations}
    \begin{align}
        2\epsilon \tan \tilde v_0 & = \epsilon_0 \sin \left( \tilde v_0 - v_0\right) \;, \\ 
        -4\sqrt{3} \sinh (2\beta) & = \epsilon_1 \sin \left( \tilde v_1 - v_1\right) \;.
    \end{align}
\end{subequations}
The first equation determines the value of $\tilde v_0$ while the second equation can be solved in the limit of a small $\epsilon_1$, which gives
\begin{equation}
    \beta \approx  \frac{\epsilon_1}{8\sqrt{3}} \sin \left[ v_1 - \tilde v_0 + \tan \tilde v_0 \left( 1 + \chi^{-\epsilon} \right)\right] + \mathcal{O}(\epsilon_1^2) \;.
\end{equation}
This shows that the complete form of a bulk axion potential would lead to a similar behavior of $\beta(\chi)$ as given in the main text with a slightly different dependence on $\tilde v_0$ and $v_1$.
The effective quartic coupling of the dilaton potential $F[\beta]$ depends on only the $\beta$ explicitly (there is no explicit $\chi$-dependence in $F$). 
Therefore, the dilaton potential is in the same form as in eq.~\eqref{eq:V_eff}.
Since $\epsilon  \approx - \tilde \epsilon \cos \tilde v_0/4$, its sign can be flipped by choosing $\tilde v_0$,
which explains the tachyonic mass term in eq.~\eqref{eq:v_b}.

\bibliography{ref}
\bibliographystyle{JHEP}

\end{document}